\documentclass[%
 reprint,
 amsmath,amssymb,
 aps,nofootinbib,   
]{revtex4-1}
\usepackage{bm}
\PassOptionsToPackage{linktocpage}{hyperref}
\usepackage[hyperindex,breaklinks]{hyperref}
\usepackage{enumitem}
\usepackage{slashed}

\renewcommand{\theta}{\vartheta}

\renewcommand{\vec}[1]{\ensuremath{\boldsymbol{#1}}}

\usepackage{array}
\usepackage{mathtools}

\usepackage{etoolbox}

\begin{document}

\title{Unitarity Entropy Bound: Solitons and Instantons}

\author{Gia Dvali} 
\affiliation{%
Arnold Sommerfeld Center, Ludwig-Maximilians-Universit\"at, Theresienstra{\ss}e 37, 80333 M\"unchen, Germany, 
}%
 \affiliation{%
Max-Planck-Institut f\"ur Physik, F\"ohringer Ring 6, 80805 M\"unchen, Germany
}%
 \affiliation{%
Center for Cosmology and Particle Physics, Department of Physics, New York University, 726 Broadway, New York, NY 10003, USA
}%

\date{\today}

\begin{abstract} 
 We show that non-perturbative entities such as solitons and instantons saturate bounds on entropy when the theory saturates unitarity. Simultaneously, 
 the entropy becomes equal to the area of the soliton/instanton. This is 
 strikingly similar to black hole entropy despite 
 absence of gravity.  
  We explain why this similarity is not an accident. 
 We present a formulation that allows
 to apply the entropy bound to instantons. The new formulation 
 also eliminates apparent violations of the Bekenstein entropy bound 
 by some otherwise-consistent unitary systems. 
 We observe that in QCD, an isolated instanton of fixed size and position violates the entropy bound for strong 't Hooft coupling.  
 At critical 't Hooft coupling the instanton entropy is equal to its area.  
\end{abstract}

\maketitle
\section{Unitarity and Entropy}

The Bekenstein entropy bound 
\cite{BekBound} (see also, \cite{BrBound})
\begin{equation} \label{Bek1}
    S_{\max} =  MR \, ,  
\end{equation}
puts an upper limit on the amount of information stored in a system of energy $M$ and size $R$.  Throughout the paper, in most of the expressions for bounds we shall drop the obvious numerical coefficients in order to focus on scaling.   We shall restore them whenever required. 
The Bekenstein bound (\ref{Bek1}) is independent of gravity.
Despite this, it attracted enormous attention in the context of gravity.  
 The reason is that in gravity the bound is saturated by black holes and  this saturation comes in form of the area \cite{BekE},
\begin{equation} \label{Bek2}
      S_{\max} = MR = (RM_P)^{d-1} \, . 
\end{equation}
Here $M$ is the mass of a black hole in $d+1$ dimensional space-time, 
$R$ is its radius and 
 $M_P$ is the  Planck mass.  \\
  
 In a recent paper \cite{Gia1}, we have posed the following questions: 
 \begin{itemize}
  \item  What is the physical meaning of saturation of 
  the Bekenstein bound 
  in non-gravitational quantum field theories? 
  
  \item  Is the area-form of saturation unique to gravity? 
   
\end{itemize} 
 We have attempted answering these questions in the following sequence. 
 First,  we showed that both 
properties (saturation of the bound as well as its area form) take place
already 
 in {\it renormalizable}  quantum field theories. 
  For example, we have observed such a behaviour for a 't Hooft-Polyakov monopole \cite{Mon} in gauge theories and for a baryon \cite{WittenN} in QCD 
  with large number of colors \cite{planar}.  We have observed that 
  these objects saturate the Bekenstein entropy bound 
 exclusively when the theory saturates the bound on unitarity. Simultaneously, the entropy acquires the form of the area. 
 Putting it shortly: 
   \begin{equation} \label{slogan}  
   {\rm Bekenstein = unitarity = area} \,.  
  \end{equation}   
 The explanation given in \cite{Gia1} to the above phenomenon
is that the key role both in
 saturation of the entropy bound as well as in its area-form is played by an 
 inter-particle coupling constant  $\alpha$ that controls perturbative unitarity of the theory. 
  The idea is that the objects that share the property (\ref{slogan}) are {\it 
  maximally packed} in the sense of \cite{Nportrait}. That is, in a certain well-defined 
 sense such an object can be viewed as a self-sustained bound-state of elementary constituents.  They have a certain characteristic 
  wavelength $R$ and a critical occupation number 
  $\sim {1 \over \alpha}$. 
 Consequently, for such systems the entropy bound can be written in the form 
in which the mass of an object is not entering explicitly but only 
the coupling constant. 
 We observe that the following features are universal.  
 
 \begin{itemize}
  \item In any consistent $d +1$-dimensional effective field theory
  the entropy $S$ of a self-sustained non-perturbative solitonic state 
   of radius $R$    
satisfies the following bound 
\begin{equation} \label{Nentropy} 
 S \leqslant {1 \over \alpha} \, , 
\end{equation}
where $\alpha$ is a relevant dimensionless
quantum coupling constant evaluated at energy scale $1/R$. 
  \item This bound is saturated when the theory saturates unitarity.  
  
  \item  There always exists a well-defined physical scale $f$ 
 that plays the role of a symmetry-breaking order parameter and 
 also sets the coupling strengths in the low energy theory.

\item The coupling $\alpha$ evaluated at the scale $1/R$ is equal 
to the inverse area measured in units of $f$: 
 \begin{equation} \label{Darea} 
 {1 \over \alpha} = (Rf)^{d-1} \,. 
\end{equation} 
This ensures the area form of the entropy at the point of saturation: 
 \begin{equation} \label{Newbound} 
 S_{\rm max} = {1 \over \alpha} = (Rf)^{d-1} \,. 
\end{equation}

\end{itemize}

 In each example,  the scale $f$ can be determined unambiguously. 
 For example, in gravity it is given by the Planck mass $f=M_P$ which sets the strength of the graviton-graviton interaction. On the other hand, for 
topological solitons such as  't Hooft-Polyakov monopole the same scale 
 is set by the vacuum expectation value of the Higgs field, $f =v$. 
 At the same time, in case of a baryon/skyrmion  $f$ is set by the pion decay 
 constant $f_{\pi}$.  
 We thus encounter the following  universal pattern at the saturation point, 
 \begin{equation} \label{Center}
  {\rm max.~entropy}  \, =\, {1 \over {\rm coupling}} = {\rm area} \,.   
  \end{equation} \\
  
  For all isolated solitons existing in Lorentzian space-times 
  the bound (\ref{Nentropy}) fully agrees with the standard Bekenstein 
  bound (\ref{Bek1}).    
  However, the form (\ref{Nentropy}) allows for a consistent generalization
  of the concept of the entropy bound to Euclidean field configurations 
  such as instantons that describe virtual processes rather than 
  states. To such entities the standard formulation of Bekenstein bound (\ref{Bek1}) cannot be applied directly since their energy is not defined.  \\

 We proceed in the following steps. 
 First, we generalize the analysis of \cite{Gia1} to solitons  in 5D theory. 
 We consider examples of scale-independent solitons in 
 theories with global and gauge symmetries. 
 The forms of these 
 solitons are identical to instantons
 in corresponding 4D theories. 
   We show that 5D solitons saturate both bounds, 
  (\ref{Bek1}) and  
   (\ref{Nentropy}), simultaneously and together with unitarity. 
  We check explicitly that at the saturation point the entropy is equal to the area, in full accordance to (\ref{Newbound}).   \\

Next,  using the formulation of the bound (\ref{Nentropy}) 
 and the connection between the 5D soliton and 4D instanton, 
 we generalize the concept of the Bekenstein entropy 
 bound to an instanton. 
 Note, we are not talking about the entropy for a  
  gas of instantons but rather {\it the entropy of a single 
 isolated instanton of  fixed size and position}.

 We assign the instanton a micro-state entropy given 
 by the entropy of its counterpart soliton in a theory with one dimension higher. 
 Assigning an entropy to an instanton may sound unusual since the instanton does not describe a ``real" object but rather a tunnelling process.  Nevertheless, using the connection between an instanton
 and a soliton in one dimension higher gives a well-defined physical meaning 
 to this assignment.  \\
  
 Next,  using the form (\ref{Nentropy}) we impose the Bekenstein bound
 on an instanton in a self-consistent way.   
  We then discover that the instanton entropy saturates the bound 
  and assumes the form of the area when the theory saturates the bound on unitarity.  This happens when the 't Hooft coupling
 assumes a critical value order-one.    
  This gives us confidence in meaningfulness of generalizing the concept of entropy to virtual states and in imposing the entropy bound on such 
  states.  \\
  
  Finally, in appendix, we provide an explicit example 
 illustrating how the formulation (\ref{Nentropy}) eliminates an apparent violation of the Bekenstein 
 bound (\ref{Bek1}) by a fully consistent unitary theory.

 \section{Scalar No-Scale Soliton in 5D}
 
 As first example we consider a scale-invariant soliton in 5D theory
 without gauge redundancy. 
 The simplest Lagrangian which leads to such a soliton has the following form, 
   \begin{eqnarray}   \label{Lsigma1} 
    &&  L_{\sigma} = {1 \over 2} \partial_{\mu}\sigma 
  \partial^{\mu}\sigma \, + \, 
  {1\over 4} g^2  \sigma^4 
    \end{eqnarray}  
  Here $\sigma$ is a real scalar field.   
  The parameter 
  $g^2 > 0$ is a five-dimensional coupling constant that has the dimensionality $[g^2] = [{\rm mass}]^{-1}$. 
The canonical dimensionality of the scalar field 
is $[\sigma] = [{\rm mass}]^{{3\over 2}}$. 
 The non-perturbative solutions of 4D version of the above theory are very well-known \cite{scalarinst}.
 Due to an attractive self-coupling the energy of the theory 
 (\ref{Lsigma1}) is unbounded from below. 
 Despite this,  it represents an useful prototype model for testing our ideas.   We shall later move to a gauge theory that does not suffer from such a problem.   \\
 
The dimensionless four-point quantum coupling at energy scale $E$ is given by 
\begin{equation} \label{alpha}
\alpha \equiv g^2E\,.
\end{equation} 
 Due to a dimensionful gage coupling the theory violates perturbative 
 unitarity above the following cutoff scale, 
 \begin{equation} \label{cutoffs}
 \Lambda \sim {1 \over g^2} \,. 
\end{equation} 
That is, above the scale $\Lambda$ the {\it dimensionless}  coupling $\alpha$ becomes strong. The above theory has a time-independent spherically symmetric solution of the following form,  
 \begin{eqnarray}   \label{sol1} 
    &&   \sigma \, = \, {\sqrt{8} \over g} { R
    \over \vec{x}^2 + R^2}\,,     
 \end{eqnarray} 
 where $\vec{x}$ is the space four-vector. 
 The form of the solution is identical to an instanton that appears in 4D version of the theory \cite{scalarinst} where it describes a tunneling process.   
 The mass of the soliton is given by 
 \begin{equation} \label{massS} 
 M_{\rm sol}  \, = \, {8 \pi^2 \over 3g^2} \,, 
 \end{equation} 
 and is independent of the localization radius $R$. Correspondingly the 
 moduli space of the soliton consists of translation and dilatation zero modes
 that can be regarded as Goldstone bosons of corresponding broken symmetries.  These modes are not sufficient for endowing the soliton 
 with a large entropy that can be increased parametrically.
In order to achieve such an entropy, we need to increase the number 
of localized zero modes. We shall accomplish this by assuming that  
$\sigma$-field transforms as a large irreducible representation of some 
internal ``flavor" symmetry group.  For example, 
$\sigma_{\alpha}, ~\alpha=1,2,...,N$ can form an $N$-dimensional representation of a 
global $SO(N)$-symmetry.  Correspondingly, the Lagrangian now takes the form, 
  \begin{eqnarray}   \label{Lsigma2} 
    &&  L_{\sigma} = {1 \over 2} \partial_{\mu}\sigma_{\alpha} 
  \partial^{\mu}\sigma_{\alpha} \, + \, 
  {1\over 4} g^2  (\sigma_{\alpha}\sigma_{\alpha})^2\,. 
    \end{eqnarray}  

Note, with the enlargement of symmetry the unitarity 
bound becomes, 
  \begin{equation} \label{unitarity1} 
  N  \lesssim {1 \over \alpha} \equiv {1 \over (Eg^2)}  \, .
   \end{equation} 
   Correspondingly, the cutoff $\Lambda$ is no longer defined by 
 (\ref{cutoffs}) but instead by the following relation,  
     \begin{equation} \label{unitaritys} 
  N  \sim  {\Lambda^{-1} \over g^2}  \, .
  \end{equation} 

The soliton solution (\ref{sol1}) goes through almost unchanged. However, it becomes highly degenerate: 
 \begin{eqnarray}   \label{solN} 
    &&   \sigma_{\alpha} \, = \, {\sqrt{8} \over g} { R
    \over \vec{x}^2 + R^2}  {a_{\alpha} \over \sqrt{N_{\sigma}}}  \,,       
 \end{eqnarray} 
where $a_{\alpha}$ are arbitrary subject to the following constraint, 
\begin{equation}\label{conN}  
 \sum_{\alpha=1}^N a_{\alpha}^2 = N_{\sigma} \,.  
  \end{equation} 
  The mass of the soliton is given by the same expression (\ref{massS}). 
  In classical theory the value of the parameter $N_{\sigma}$ 
  is irrelevant. However, in quantum theory it acquires an important meaning,
  as we shall explain. \\ 
  
  The above degeneracy can be understood in two equivalent languages. 
 The first is the language of a moduli space.  Since the soliton became embedded 
 into the $SO(N)$-space,  its moduli space got increased. 
 Indeed, an arbitrary orthogonal $SO(N)$-transformation,
 \begin{equation} \label{Utrans} 
 \sigma_{\alpha} \rightarrow O_{\alpha\beta}\sigma_{\beta}\,, 
 \end{equation} 
 that acts on the soliton non-trivially, gives again a valid solution
 with exactly the same mass $M_{\rm sol}$ and size  $R$. 
  Thus, the soliton's  internal moduli space becomes $SO(N)/SO(N-1)$ 
  which has a topology of an 
 $S_{N-1}$-sphere.  Thus, we have $N-1$ moduli parameterizing 
 the location of the soliton on this $N-1$-dimensional sphere. \\ 
  
  An alternative language for describing  the soliton degeneracy  
   is of Goldstone modes. The non-zero 
 expectation value of the $\sigma$-field breaks spontaneously the global $SO(N)$-symmetry 
 down to $SO(N-1)$. Consequently, there emerge $N-1$ Goldstone modes localized within the soliton. 
  \\
 
  The situation that we got is fully analogous to the example of a
  't Hooft-Polyakov monopole constructed in \cite{Gia1}. There
  too, a global $SO(N)$-symmetry was spontaneously broken within the monopole core.  Therefore, the counting of the micro-state degeneracy goes in the same way. Of course, classically, the degeneracy of the 
  moduli space is infinite, since every point on it counts as a different
  Goldstone vacuum of the soliton. This is also a manifestation of the fact that in $\hbar = 0$ limit the entropy 
  is infinite, as it should be.  Of course, in quantum theory the entropy of a localized soliton becomes finite.  \\

  The micro-state degeneracy can be deduced from the 
  effective Hamiltonian describing the vacuum structure of the soliton.
  This can be written in the following simple form \cite{Gia1},  
  \begin{equation}\label{gold}  
  \hat{H}  = X \left(\sum_{\alpha = 1}^N \hat{a}_{\alpha}^{\dagger} \hat{a}_{\alpha} - N_{\sigma}\right ) \, ,   
  \end{equation} 
  where $X$ is a Lagrange multiplier which enforces  the constraint 
  (\ref{conN}) and the parameter $N_{\sigma}$ will be determined below.  
 All the non-zero frequency modes have been excluded as they 
 do not contribute into the ground-state structure but only into the excited states. 
 Therefore, the operators       
$\hat{a}_{\beta}^{\dagger}, \hat{a}_{\alpha}$ represent the creation and annihilation operators of the {\it zero frequency}  moduli.  They satisfy the
 usual commutation relations  
$[\hat{a}_{\alpha}, \hat{a}_{\beta}^{\dagger}] = \delta_{\alpha\beta}$.
Obviously, the quantities $a_{\alpha}^2$ entering in (\ref{conN}) 
  and (\ref{solN}) must be understood as the expectation values 
  of the number operators 
  $\hat{n}_{\alpha} \equiv \hat{a}_{\alpha}^{\dagger}\hat{a}_{\alpha}$. 
  The micro-state degeneracy is then given by all possible 
  distributions of these numbers subject to the constraint 
  (\ref{conN}).   
 The total number of such states is given by
  the binomial coefficient 
  \begin{equation}\label{Nstates} 
 n_{\rm st} =  \begin{pmatrix}
    N_{\sigma} + N -1   \\
     N_{\sigma}   
\end{pmatrix} \,. 
  \end{equation}
  In order to evaluate this expression, we need to determine the value of the parameter $N_{\sigma}$ which of course must depend on $R$.  
 For fixing it, notice that we can describe the soliton 
 as the bound-state of $N_{\sigma}$ quanta of the field $\sigma$ 
 and treat it in Hartree approximation essentially following the method applied by Witten to baryons/skyrmions in \cite{WittenN}.
\footnote{Here we use the bound-state view for a simple estimate, otherwise not relying on it.  More detailed steps in understanding corpuscular 
structure of solitons/instantons were taken in \cite{corpuscular}.}.
 Indeed, the bound-state has the size $R$ and is stabilized  due to an attractive interaction measured by the coupling $g^2$. 
  This means that the positive kinetic energy of each quantum, which scales as $\sim 1/R$, is exactly balanced by the negative potential energy of attraction  
from the rest, which scales as $\sim g^2N_{\sigma}$. We thus 
  conclude, 
   \begin{equation}\label{Nsigma}  
  N_{\sigma} \sim  {R \over g^2}  \, .
   \end{equation}

 Now, comparing the above expression to (\ref{unitarity1}), it is clear that whenever $N$ saturates the unitarity 
 bound at energy $1/R$, we have  
 \begin{equation} \label{2N}
 {\rm unitarity~limit}  \rightarrow  N_{\sigma} \sim N\,.
 \end{equation} 

 Or equivalently, this equality takes place whenever the soliton size becomes equal to the unitarity 
 cutoff scale $R = 1/\Lambda$.  In this case, 
 from (\ref{Nstates}) using Stirling's approximation we get that  
the micro-state entropy of the soliton in the unitarity limit becomes equal to, 
  \begin{equation}\label{Ssoliton}  
  S_{sol}  = \ln(n_{\rm st}) \sim N  \, .
   \end{equation} 
   
  We are now ready to test our claims. The first task is to see explicitly that the above entropy saturates the Bekenstein bound (\ref{Bek1}).  
Next, we wish to see that this saturation is in full agreement with (\ref{Nentropy}). 
 Thirdly, we must show that at the saturation point the entropy assumes the form of the area according to (\ref{Newbound}).  \\
 
 Taking into account the expression for the soliton mass (\ref{massS}), the Bekenstein bound (\ref{Bek1}) reads, 
    \begin{equation}\label{Solmax} 
 S_{\rm max}  \sim  {R \over g^2} \, .
  \end{equation}
 Notice, this is exactly the bound (\ref{Nentropy}) with 
 $\alpha = g^2E$ evaluated at the scale $E=1/R$.   
Thus, the two bounds fully agree. 
They become saturated when (\ref{Solmax}) becomes equal to 
(\ref{Ssoliton}). This happens when 
    \begin{equation}\label{Ssat} 
 N  \sim {R \over g^2} \, .
  \end{equation}
   As it was already said, by taking into account (\ref{unitarity1}), 
the expression (\ref{Ssat}) means that  $N$ saturates the unitarity bound 
 and  at the same time $R = \Lambda^{-1}$.  
 Indeed, for $R = \Lambda^{-1}$ the equation (\ref{Ssat}) takes the form 
  (\ref{unitarity1}).    
   Thus, we see that the soliton saturates the Bekenstein bound (\ref{Bek1}) (and simultaneously (\ref{Nentropy})) 
  precisely when its size $R$ becomes 
  equal to the cutoff scale $\Lambda^{-1}$ fixed by the unitarity bound (\ref{unitarity1}). 
   \\
  
   Let us now investigate whether at the saturation point the entropy  takes the form of the area (\ref{Newbound}).  For this, we first need to identify the scale $f$. 
   This scale is determined by the order parameter that breaks 
   the $SO(N)$-symmetry spontaneously. 
  The latter is given by the maximal value that the 
 field $\sigma$ reaches in the core of the soliton
 and is equal to
 \begin{equation}  \label{simacore}
 \sigma_{\rm max} = \sigma(0) =  {\sqrt{8} \over gR}\,.
 \end{equation}
   Since, $\sigma$ has dimensionality 
 of $[{\rm mass}]^{{3\over 2}}$, it is clear that the relevant scale in the problem is 
   \begin{equation}\label{scalef}   
 f =  {1 \over (gR)^{{2\over 3}}}\,.  
 \end{equation}   
   This is the  correct scale that must be used 
   for measuring the surface area of the soliton. 
   Expressing then  (\ref{Ssat}) through (\ref{scalef}) it is clear   that the entropy at the saturation point can be written as the surface area in units of $f$: 
      \begin{equation}\label{Area5S} 
 S_{\rm sol}  \sim (R f)^3 \, .
  \end{equation}
  Note, the surface area of a soliton in 5D is a three-dimensional surface.  So, (\ref{Area5S})  is in full accordance with (\ref{Newbound}). \\ 
    
     In summary, we  observe that an instanton-like soliton  
  in 5D exhibits exactly the same tendency as was observed earlier for 
  't Hooft-Polyakov monopoles and baryons in \cite{Gia1};   the saturation of the Bekenstein bound (\ref{Bek1}) is equivalent to the saturation of the bound (\ref{Nentropy}) and both are saturated together with the unitarity
  bound (\ref{unitaritys}).  The corresponding entropy exhibits the 
  area-law (\ref{Area5S}) in agreement with the general 
  relation (\ref{Newbound}).

 \section{No-scale Gauge Monopole} 
 We shall now discuss a gauge soliton in 5D.
 We first consider a simplest model that contains such a soliton. 
  This is a theory with a gauged $SO(3)$ symmetry
with a triplet of gauge fields  $A_{\mu}^a$ where $a = 1,2,3$ is
an $SO(3)$-index.  Sine we are in five space-time dimensions, 
the indexes $\mu,\nu$ take values $0,1,2,3,4$.
    The Lagrangian
   has the following form:  
     \begin{eqnarray}   \label{Lag1} 
  &&  L =  
    - {1 \over 4} F_{\mu\nu}^aF^{\mu\nu a}
   \end{eqnarray}
 where $F_{\mu\nu}^a \equiv \partial_{\mu}A_{\nu} ^a -\partial_{\nu}A_{\mu} ^a
  + g \epsilon^{abc} A_{\mu}^bA_{\nu}^c$. The parameter 
  $g$ is a five-dimensional gauge coupling constant that has a dimensionality $[g] = [{\rm mass}]^{-{1\over 2}}$. 
The canonical dimensionality of five-dimensional gauge field 
is $[A_{\mu}^a] = [{\rm mass}]^{{3\over 2}}$. 
The dimensionless four-point coupling at energy scale $E$ is 
given by the same expression (\ref{alpha}) as in the previous example.  
 The same is true about the cutoff scale $\Lambda$  above which the coupling  $\alpha$ becomes strong and violates unitarity. 
 It is given by  (\ref{cutoffs}). \\

  The above theory admits a time-independent localized monopole-like 
soliton solution with the topological charge given by  
$Q= {g^2 \over 32 \pi^2} \int d^4x \epsilon^{0\mu\nu\alpha\beta} F_{\mu\nu}^aF_{\alpha\beta}^a$. 
This soliton is identical to 
 an instanton solution in four-dimensional Euclidean space
 \cite{instanton}. Therefore, when lifted to a five-dimensional space-time with the Lorentzian  
 signature, it describes a localized monopole-like soliton.
 These 5D solitons and their relation to 4D instantons  are well known 
 and were studied previously in various contexts, in particular, 
 in the context of large extra dimensions \cite{Hill}.

 For definiteness, we shall focus 
on the case $Q =1$. The solution has the well-known form: 
  \begin{eqnarray}   \label{monopole} 
    &&   A_{\mu}^a \, = \, {2 \over g} \eta^a_{\mu\nu} {x^{\nu} 
    \over \vec{x}^2 + R^2}\,,    \\ \nonumber
   &&  F_{\mu\nu}^a \, = \, -\, {4 \over g} \eta^a_{\mu\nu} {R^2 
    \over (\vec{x}^2 + R^2)^2}  \, ,    
 \end{eqnarray} 
 where $\eta^a_{\mu\nu}$ are 't Hooft's  parameters that shall not be displayed explicitly.  As in the case of the soliton in previous example, the mass of the gauge monopole  is independent of $R$ and is 
 given by the similar expression, 
 \begin{equation} \label{massM} 
 M_{\rm mon}  \, = \, {8 \pi^2 \over g^2} \,. 
 \end{equation} 
 
  The zero mode spectrum of the above monopole consists of a standard set
 of translation, dilatation and orientation moduli, eight in total. 
 These are not sufficient for 
 delivering a micro-state entropy that we could increase in a controllable way.  For achieving the latter goal, we must increase the number of 
 bosonic and/or fermionic gapless modes localized within the monopole as it was done in \cite{Gia1}.  
 
 \section{Bosonic zero modes} 
 
   In this case we need to enlarge the monopole moduli space 
   by embedding the $SU(2)$ gauge symmetry as a subgroup into a larger
 $SU(N)$ symmetry group.   This embedding does not change the expression for the monopole mass (\ref{massM}). However, the unitarity 
 constraint becomes (\ref{unitarity1}) with the unitarity cutoff
 scale $\Lambda$ given by (\ref{unitaritys}).   
In order to fix the numerical coefficients, we shall define a five-dimensional analog of the 't Hooft coupling in the following way,
 \begin{equation}  \label{thooft}
 \lambda_{t}  \equiv N {g^2 E\over 8\pi^2} \,. 
 \end{equation}
 Obviously, for strong $\lambda_t$ the theory violates perturbative unitarity.  \\

  The above embedding results into the appearance of additional 
  bosonic zero modes. These modes parameterize the orientation 
  of the monopole's $SU(2)$-subgroup within $SU(N)$. 
  They correspond to global $SU(N)$-transformations that act non-trivially 
  on the monopole.  
  Therefore, the orientation moduli space is  
 $SU(N)/SU(N-2)\times U(1)$.
 However, the $U(1)$-factor is only partially residing within the stability group 
 of the monopole.  
 At large $N$ this moduli-space has dimensionality $\sim 4N$.   This space 
 is  identical to a standard moduli space 
 of an instanton of $SU(N)$ gauge theory in 4D. The 
 novelty here is that we are interpreting this degeneracy in terms of the Goldstone  phenomenon and giving it a meaning of the micro-state entropy. 
 Next, we are correlating the resulting entropy bound with the unitarity and the area of the system. \\
 
  Indeed, the moduli space can be understood as the degeneracy of the 
  soliton vacuum due to spontaneous breaking of $SU(N)$ global 
  symmetry by the monopole.  This is fully analogous 
  to the breaking of global $SO(N)$ symmetry by a soliton considered in the previous example.  
 Consequently,  up to order-one factors, the counting of the entropy is 
 very similar. Thus, the micro-state degeneracy is again described by the effective 
 Hamiltonian of the form (\ref{gold}) with the number of zero modes 
 now being of order $4N$
  \begin{equation}\label{gold}  
  \hat{H}  = X \left(\sum_{\alpha = 1}^{4N} \hat{a}_{\alpha}^{\dagger} \hat{a}_{\alpha} - N_{\rm mon}\right ) \, ,   
  \end{equation} 
 and the parameter $N_{\rm mon}$ 
 given by (\ref{Nsigma}).  However, we would like to be more precise. 
 We shall make a guess and choose $N_{\rm mon}$  to be a dimensionless quantity
 constructed out of the size of the soliton $R$ and the topological invariant
 \begin{equation} \label{Nmon}
 N_{\rm mon} = R \int d^4x \epsilon^{0\mu\nu\alpha\beta} F_{\mu\nu}^aF_{\alpha\beta}^a
  = R {32 \pi^2 \over g^2 } \, .
  \end{equation}  
 The number of resulting micro-states at large-$N$ is given by the expression analogous 
 to (\ref{Nstates}), 
  \begin{equation}\label{NstatesMon} 
 n_{\rm st} \simeq  \begin{pmatrix}
    N_{\rm mon} + 4N   \\
     N_{\rm mon}   
\end{pmatrix} \,. 
  \end{equation}
Interestingly, this qualitatively matches what would 
be the pre-factor measure for the contribution of  4D instanton of size $R$ into 
the vacuum-vacuum transition probability. In particular, for  weak 't Hooft coupling the equation (\ref{NstatesMon}) can be written as 
  \begin{equation} \label{Nmon}
 n_{\rm st} \sim  
 { 1 \over (N!)^4} \left ({8 \pi^2 R\over g^2} {\rm e}^{\lambda_t}\right )^{4N}  \, .
  \end{equation}  
  The visual similarity with the pre-factor in a standard 
  instanton transition probability \cite{prefactor, review} is clear. 
  The underlying physical reason for this connection shall become more transparent below, after we extend the notion of entropy to instantons. 
  \\

 The entropy of the monopole/instanton  saturates the Bekenstein bound 
 when the 't Hooft coupling becomes order one. 
 In order to make this more precise, let us evaluate the monopole entropy for $\lambda_t =1$. 
First,  taking into account (\ref{thooft}) and 
 (\ref{Nmon}) we see that in this limit $N_{\rm mon} = 4N = R {32 \pi^2 \over g^2 }$.  Next, plugging this in (\ref{NstatesMon}) and using Stirling's 
 approximation, we find that the monopole entropy 
 for $\lambda_t =1$ takes the following form  
 \begin{eqnarray}
 \label{MonBound} 
 S_{\rm mon} &=& \ln(n_{\rm st}) \simeq {64 \pi^2 \over g^2}R \ln(2) \simeq
 \\ \nonumber 
 &\simeq&  1.3 \left( {4\pi \over 3} M_{\rm mon} R \right ) \, , 
  \end{eqnarray} 
 where $M_{\rm mon}$ is given by (\ref{massM}).  
 The last expression in the brackets 
 is nothing but a 5D Bekenstein bound on an entropy of an object of mass $M_{\rm mon}$ 
and the size $R$,  
\begin{equation} \label{SBH}
S_{\max} = {4\pi \over 3} M_{\rm mon} R\,.  
\end{equation}
At the same time, of course, this expression is equal to the
Bekenstein-Hawking entropy  of a  would-be static black hole 
with the same parameters but in 5D theory with gravity. 
\\

 We thus observe that the monopole/instanton entropy saturates 
 the Bekenstein bound when 't Hooft coupling is close to one.
 Note, since the sensitivity to the value of $\lambda_t$ is exponential, 
the pre-factor $1.3$ simply indicates that the saturation takes place 
when $\lambda_t$ is very close to one. 
 Equivalently,  we can write,  
\begin{equation} \label{ALMOST} 
 S_{\rm mon}(\lambda_t \simeq1) =  S_{\max}\,.  
  \end{equation}

 Simultaneously, the entropy acquires a form of the area measured in units of the scale  (\ref{Area5S}) with the scale $f$ given by (\ref{scalef}). 
 Notice, the scale $f$ has a very well defined physical meaning, 
as it represents an order parameter of spontaneous breaking of the global 
$SU(N)$-symmetry by the monopole. 
Equivalently, it sets the interaction strength of the orientation moduli.  
In this sense, there is no ambiguity in defining the physical meaning 
of the scale $f$. 
Thus, the area form of the entropy at the saturation point is unambiguous. \\

 It is interesting to map the entropy of a monopole 
 on an entropy of a wold-be black hole of the same mass and the size. 
   For this, we write the monopole entropy in  form of the black hole entropy by explicitly separating $1/4$: 
 \begin{equation} \label{1over4}  
   S_{\rm mon} = {{\mathcal A} \over 4 G}  \,, 
 \end{equation}
 where ${\mathcal A} \equiv 2\pi^2R^3$ is the monopole surface area
 in 5D, and the parameter $G$ has the following form, 
  \begin{equation} \label{1over4}  
   G \equiv {3 \over 64\pi} g^2R^2 \equiv f^{-3}\, .    
 \end{equation}
  As said above, $G$ has a very transparent physical meaning
  as it sets the coupling strength of orientation moduli that represent the Goldstone bosons of broken global symmetry. In this sense it is fully analogous to the Newton's constant that sets the interaction strength among gravitons.  Correspondingly, the scale $f$ plays the role of the Planck mass.

  \section{Fermion zero modes}

  An alternative possibility for increasing the micro-state entropy is to  populate the 5D monopole by fermion zero modes, as it was done 
in \cite{Gia1}. 
This is accomplished via coupling the gauge field to a large number of fermion species. 
Let us assume that fermions form an $N$-dimensional representation of 
  a global flavor symmetry group. 
  Due to existence of $\sim N$ fermion species the unitarity bound 
  is given by (\ref{unitarity1}) as in the previous case. 
  Correspondingly, the cutoff scale $\Lambda$  is determined by (\ref{unitaritys}). \\

    As it is well-known, in the instanton/monopole background fermions  
    give rise to localized zero modes. 
     A detailed construction of fermion zero modes in the instanton backgrounds can be found
 in several excellent reviews \cite{review}. For us, the important fact is that the number of zero modes scales as $\sim N$ and can be made arbitrarily large
 by increasing the number of flavors. 
 \\
   
 For any given $R$,  the  existence of $N$ fermionic zero modes 
 creates  
 $n_{\rm st} = 2^{N}$ degenerate micro-states in the monopole spectrum.
 Correspondingly, the monopole acquires a  micro-state entropy 
give by,  
   \begin{equation}\label{SmonF} 
 S_{\rm mon} = \ln (n_{\rm st}) \sim N \, .
  \end{equation} 
  From here, it is obvious that all the effects observed in the previous 
  examples repeat themselves. This can be easily seen by taking 
  into account (\ref{unitarity1}) and (\ref{unitaritys}) and 
  comparing (\ref{SmonF}) with (\ref{Solmax}).  
  Thus again the monopole saturates the entropy bounds (\ref{Bek1}) 
  and (\ref{Nentropy}) together with unitarity.  
  The area form of the entropy 
  (\ref{Area5S}) at the saturation point also remains intact.

  \section{Instanton in 4D} 
  
    We shall now move to 4D and consider an instanton instead of a soliton.  
    It is given by exactly the same solution (\ref{monopole}) but 
    ``downgraded" to 4D Euclidean space.   Therefore, we have to 
  take into account the change of  dimensionalities of the parameters.  We shall denote them by the same symbols as in 5D. However, we must remember that $g$ now is dimensionless and $A_{\mu}^{a}$   has a dimensionality of mass.  Correspondingly, we define the 
four-dimensional analog of the 't Hooft coupling as
 \begin{equation}  \label{thooft4}
 \lambda_{t}  \equiv N {g^2 \over 8\pi^2} \,.  
 \end{equation}
Note, in 4D we wish to keep the theory asymptotically-free.
  So we must assume that the fermion content is chosen appropriately. 
  In the present case, for simplicity,  we shall ignore the fermion flavors. \\

    We shall now undertake the following two steps.   
First, we shall assign entropy to an instanton. 
Secondly, we shall try to understand what is the upper bound on
this entropy and what is the physical meaning of its saturation.  \\   
    
    We assign the entropy to the instanton by a direct generalization of the 
entropy of its 5D counterpart soliton.      
  Namely, the entropy of the instanton will be counted  as the 
 log of the number of micro-states (\ref{Nmon}) due to the existence 
 of zero mode bosons and/or fermions,  exactly as it was counted for the 5D monopole. So we shall assign to an instanton the following number of 
 ``micro-states", 
  \begin{equation}\label{NstatesInst} 
 n_{\rm st} \simeq  \begin{pmatrix}
    N_{\rm inst} + 4N   \\
     N_{\rm inst}   
\end{pmatrix} \,,  
  \end{equation}
  where taking into account the dimensionality of the coupling 
  constant we take, 
\begin{equation} \label{Ninst}
 N_{\rm inst} = \int d^4x \epsilon^{\mu\nu\alpha\beta} F_{\mu\nu}^aF_{\alpha\beta}^a
  = {32 \pi^2 \over g^2 } \, .
  \end{equation} 
  From the perspective of Lorentzian space, (\ref{NstatesInst}) counts 
  the number of {\it micro-processes} that cost the same Euclidean action
  $I_{inst} = {8 \pi^2 \over g^2}$.    
 For weak 't Hooft coupling the ``degeneracy" (\ref{NstatesInst}) 
can be approximated as,  
  \begin{equation} \label{Nmon}
 n_{\rm st} \sim  
 { 1 \over (N!)^4} \left ({8 \pi^2 \over g^2} {\rm e}^{\lambda_t}\right )^{4N}  \, .
  \end{equation} 
  The coupling $g^2$ is evaluated at the scale $R$.   
 This matches the expected measure for the contribution of a given size instanton into 
the vacuum-vacuum transition probability \cite{prefactor, review}.
This is remarkable since the correlation 
between the degeneracy of actual states and their contribution into the virtual processes is rather non-trivial. The interpretation of the transition probabilities in terms of instanton 
entropy can be highly instructive. Namely, the question that we would like to ask 
is: \\

 {\it What is the physical significance of the violation of the entropy bound
 by an instanton?}   \\
 
   In order to answer this question, we first need to solve the following dilemma.  
    Since we are in Euclidean space, there exists no notion of 
 energy.  So the Bekenstein entropy bound 
 (\ref{Bek1}) is not well-defined and cannot be applied to 
 an instanton directly.   
 However, the bound (\ref{Nentropy}) is well-defined and can be applied. 
 This bound is saturated when the 't Hooft coupling
 (\ref{thooft4}) becomes 
 order-one.  Indeed, taking $\lambda_t =1$ we get
  \begin{equation} \label{InstBound} 
 S_{\rm inst} = \ln(n_{\rm st}) \simeq {64 \pi^2 \over g^2} \ln(2)\,,
  \end{equation} 
where $g^2$ has to be evaluated at the energy scale $1/R$. \\

Equivalently, we can say that an instanton of a given size $R$ saturates 
the entropy bound (\ref{Nentropy}) when the counterpart monopole 
in 5D saturates the ordinary Bekenstein bound as described by 
(\ref{MonBound}).
The fact that the entropy  saturation takes place when 't Hooft 
coupling becomes order one, cannot be a simple coincidence 
and must be revealing some deep underlying physics. \\

Similarly to the 5D monopole, at the saturation point the entropy takes the 
form of the area. 
 \begin{equation} \label{1over44D}  
   S_{\rm mon} = {{\mathcal A} \over 4 G}   
 \end{equation}
 where ${\mathcal A} \equiv 4\pi R^2$ is the area
 and the parameter $G$ has the form, 
  \begin{equation} \label{G4D}  
   G \equiv {g^2R^2 \over 64\pi \ln(2) } \equiv f^{-2}\, .    
 \end{equation}
  As in 5D case, the parameter $G$ has a very well defined physical meaning as it sets the coupling strength of the Goldstone modes. 
  In this respect it is fully analogous to Newton's constant in gravity. \\
  
Notice, we get the area law as for a localized object in  4D Minkowski space-time. 
The physical meaning of this is easy to understand.
 An instanton describes a tunnelling process.  We can connect this process 
 to 5D monopole discussed in the previous section.
For this,  we can imagine that the 4D Minkowski gauge theory 
resides  on a 4D slice of the 5D gauge theory.  Now, the 5D theory houses monopoles. These monopoles can tunnel 
through the 4D surface. This ``passing-by" virtual monopole is 
``seen" by a 4D observer as  instanton. \footnote{In the present discussion 
the embedding of 4D theory into 5D is just a mental exercise
for understanding the connection, with no real physical meaning 
implied for the fifth dimension. The embedding however can be given a direct physical meaning 
in specific constructions such as \cite{Hill} or \cite{DNT}.}  
 
In such a picture, an each 
passing-through event can be attributed a particular location in 4D space. 
Of course, by 4D Poincare invariance we effectively integrate over all possible locations and sizes whenever we compute instanton contribution
into the physical observables. However, this is not important for the present discussion since we are interested in the entropy of an instanton with 
fixed size and location.  Then, the area of a sphere surrounding each event is a two-dimensional surface.

\section{Black Holes} 

Although in the present paper we focus on non-gravitational theories,
we must comment on an obvious connection with the black hole entropy
(\ref{Bek2}). 
As it was already noticed in \cite{Nportrait} the entropy of a black hole 
of size $R$ can be written in the form (\ref{Newbound}) where $\alpha$  
must be understood as the gravitational coupling at the scale $1/R$: 
\begin{equation}\label{BH}  
 \alpha_{gr} = {1 \over (RM_P)^{d-1}} \, .
 \end{equation} 
 This fact suggests that the remarkable similarity  
 between solitons/instantons/baryons/skyrmions 
 on one hand and black holes on the other  exhibited at 
 the saturation point  lies in the fact that all these 
 seemingly-different entities are {\it maximally packed} 
 composite objects \cite{Nportrait}.   
Such objects consist of maximal occupation number of quanta 
compatible with the strength of the coupling $\sim 1/\alpha$.  
It is then natural that the entropy capacity of such objects 
fully saturates  the unitarity bound of the system for a given $\alpha$.  \\

It is also clear why the Planck mass plays the role analogous to symmetry breaking parameter. Notice, the canonically normalized graviton field 
reaches value $\sim M_P$ near the black hole horizon as seen by 
an external observer in Schwarzschild coordinates. 
This is analogous to the field $\sigma$ reaching the values set by the scale 
$f$. The same is true about the baryon that can be viewed as the skyrmion 
 soliton \cite{skyrme,WittenS}. There too the maximal value of the order 
 parameter is set by the pion decay constant $f_{\pi}$.  \\
 
 As already pointed out in \cite{Gia1},  yet another similarity in 
 black hole case is the existence of the species length-scale 
 $\Lambda^{-1} = N^{1 \over {d-2}}/M_P$ \cite{speciesG}.  This is the 
 size of a smallest black hole which saturates the 
 bounds (\ref{Bek1}) and (\ref{Nentropy})  on information storage capacity. At the same time,  at the same scale, the gravitational interaction saturates unitarity.

\section{discussions} 

In the present paper we gained an additional support for 
the results obtained in \cite{Gia1} and made some further steps. 
  First, we continue to observe that manifestations of the 
Bekenstein entropy bound  that  are usually
attributed to gravity are universal. They are shared by non-perturbative objects in generic consistent theories regardless of 
renormalizability and/or the presence of gravity. 
 We observe that the saturation of the entropy bound and its area form are linked with the saturation of unitarity. \\

 For soliton- and instanton-like non-perturbative objects the entropy bound can be 
 formulated in the form (\ref{Nentropy})
in which the mass of the object does not enter explicitly.  
Instead, the bound is set by the inverse of the coupling constant. This form makes 
 the connection between the saturation of the entropy bound 
and unitarity more transparent. Through the relation (\ref{Newbound}) 
it also gives a quantum field theoretic explanation to the area-form
of the entropy at the saturation point.  \\

In appendix we highlight the importance of the coupling constant in formulation of the entropy bound (\ref{Nentropy}). Namely, we provide an explicit example of manifestly unitary theory in which the Bekenstein bound in its original formulation (\ref{Bek1}) is seemingly violated, however the formulation (\ref{Nentropy}) restores the consistency.  \\

Another novelty is that we have generalized the concept of the entropy
bound to an instanton of a fixed size and location.  
For this, we first needed to assign the entropy to an instanton.  
 We did this in two equivalent ways. First, we assign to an instanton  
an entropy that would be carried by its counterpart soliton in a space with one dimension higher.   
An alternative way is to use the formulation (\ref{Nentropy}) which for  solitons and baryons fully agrees with Bekenstein bound
(\ref{Bek1}). 
However, since the bound (\ref{Nentropy}) does not include any dependence on the mass, we  can directly apply it 
to an instanton.  We then observe that the instanton saturates the bound (\ref{Nentropy}) 
when its counterpart soliton in a theory one dimension higher 
saturates the Bekenstein bound (\ref{Bek1}). Both saturations are synchronized
with the saturations of unitarity in respective theories and take the forms of the respective areas.  The saturation of the entropy bound 
takes place when the 't Hooft coupling reaches the critical value 
order one. \\ 

It would be interesting to understand if this criticality is somehow related with possible phase transitions in large-$N$  QCD, for example,  of the type suggested by Gross and Witten \cite{GW}  and by 
Wadia \cite{Wadia}, or  with other non-analiticities at large-$N$ 
\cite{Neu}. \\

One lesson we are learning is that the solitons and instantons 
are no less holographic \cite{Hol1} than black holes.  
Our observations indicate that saturation of the entropy bound and its 
area-form go well beyond gravity and are defined by fundamental 
aspects of quantum field theory such as unitarity and asymptotic freedom.  \\

\section*{Acknowledgements}
We thank Cesar Gomez, Andrei Kovtun, Oto Sakhelashvili, 
Goran Senjanovic, 
Nico Wintergerts  and Sebastian Zell for discussions. 
This work was supported in part by the Humboldt Foundation under Humboldt Professorship Award, by the Deutsche Forschungsgemeinschaft (DFG, German Research Foundation) under Germany's Excellence Strategy - EXC-2111 - 390814868,
and Germany's Excellence Strategy  under Excellence Cluster Origins. 


\section{Appendix: Coupling and Entropy Bound}

  Here we would like to discuss the importance of the coupling 
  for formulation of the entropy bound.  This should provide an additional justification for writing the bound in the form (\ref{Nentropy}).  In order to do this, it is
 the simplest to give an explicit example of a consistent 
 field theory that {\it naively}  violates 
 the entropy bound formulated in the original Bekenstein form (\ref{Bek1}) but the violation is avoided 
  after we take into account the formulation (\ref{Nentropy}).  \\
  
   We note that apparent violations of the Bekenstein bound in gravitational context and attempts of resolving such violations where discussed previously,  e.g., in \cite{species},\cite{B1}.  
The formulation (\ref{Nentropy}) can be  useful in this respect also, as it avoids violation of the bound in any unitary theory.  \\

  Consider any unitary theory that 
  admits a large number of non-interacting localized solitons.  For example,  let us focus on a set of 
 $N$-copies  of $SO(3)_j$  gauge theories  
  in 4D, where $j=1,2,...N$ is the label of the group. 
  Assume that each copy is Higgsed by an accompanying triplet scalar field
  $\Phi_j^{a_j}$, where $a_j =1,2,3$ is the gauge index of the respective
  $SO(3)_j$.  In this way, each  $SO(3)_j$ is Higgsed down to 
  an $U(1)_j$-subgroup.  As a result, we obtain a 't Hooft-Polyakov  
 magnetic monopole residing in each theory.  For simplicity, let us assume that 
 the parameters of the theories such as gauge couplings
$e_j=e$  and masses
 are equal and the theories do not 
 talk to each other. That is, the Higgs vacuum expectation values 
 are all equal $\langle \Phi_j\Phi_j \rangle = v$ and so are the gauge boson 
 masses $m_j = ev =m_v$.  \\

 Then, we get the following situation. 
The masses of the monopole species $M_j = m_v/e^2 $ as well as their sizes 
$R_j= 1/m_v$ are all equal to each other, $M_j = M_{\rm mon}\,,  R_j =R$.   
 Since the theories are decoupled, the monopoles from different sectors 
experience no interaction and they can be placed at arbitrary distances from each other. In particular, we can place them right on top of one another and create a stack of $N$ monopoles.   The size of the stack is $R$ but the mass
scales as $M= N M_{\rm mon}$.  Thus, the maximal entropy permitted 
by the Bekenstein bound (\ref{Bek1}) scales linearly with $N$: 
$S_{\rm max} = N (MR)$.  
On the other hand the {\it naive} entropy scales as 
$S_{\rm mon}  \sim N^2$ due to the number of moduli that 
 parameterize the relative positions. Their number is growing as $\sim N^2$. So, it appears that by taking $N$ sufficiently large 
the entropy of the monopole stack can violate the Bekenstein bound (\ref{Bek1}). \\
 
 However,  the issue is more subtle. 
 Although the entropy formally grows unbounded with $N$, the information content stored in such micro-states is ``sterile" and therefore meaningless. 
 Indeed, in order to read-out the quantum information stored 
 in a state of displacement moduli, a device must exist that can measure 
 such displacements. However, in quantum field theory there exist no external devices. All the ``devices" are manufactured out of the quantum fields. 
 Thus, in order to measure the quantum information stored in the monopole
 displacements, an agent is required in form of a quantum field that interacts 
 with all the monopole species. If such an agent is absent, the information 
 is unreadable. The Bekenstein bound formulated in the form 
 (\ref{Bek1}) cannot capture this difference explicitly, since  it is independent 
 of the strength of the coupling. \\ 
 
 On the other hand, the bound  formulated as (\ref{Nentropy}) does capture the difference.  Indeed, in the absence of any mediator quantum
 field, the inter-monopole coupling
 strength $\alpha_{ij}$ vanishes. So written in the form (\ref{Nentropy}) 
 \begin{equation} \label{coupling} 
   S_{\rm max} < {1 \over \alpha_{ij}}
 \end{equation} 
 the bound is maintained for arbitrary large $N$ and consistency is restored. 
 The advantage of the bound (\ref{Nentropy}) is that it accounts for the fact that in order to read-out the information and/or to maintain the bound-state the coupling must be non-zero.  \\
 
 Let us now create a measuring device. 
 The role of it can be played by a field that couples to all the Higgs 
 triplets. For example, a gauge-singlet scalar $\chi$  and/or a Majorana gauge-singlet fermion 
 $\psi$ with the following couplings, 
 \begin{equation} \label{connector} 
   \sum _j \, ~g_j^2 |\chi|^2 (\Phi^{a_j}_j\Phi^{a_j}_j) \, 
   + \, \tilde{g}_j \Phi^{a_j}_j\bar{\psi}^{a_j}_j \psi \,,    
 \end{equation} 
 would do the job. Notice, in order to be connected via a gauge-singlet 
 fermion $\psi$, an each $j$-sector must also contain a gauge-triplet fermion
 $\psi^{a_j}$.  There is no need for this in the scalar case. \\ 
 
 For simplicity, we can assume all the $g_j$ and $\tilde{g}_j$ to be of the same order
 $g_j \sim \tilde{g}_j \sim g$. Then, unitarity puts the following bound on 
 this coupling-strength 
  \begin{equation} \label{Gbound} 
  \sum_j g_j^2 \sim g^2N   \lesssim 1 \, .
  \end{equation} 
 Now, the coupling to the connectors induces the coupling 
 among the different monopole species already at one-loop level due to 
 the exchanges by $\chi$ and $\psi$. This can be seen from the following part of one-loop effective 
 Coleman-Weinberg potential, 
 \begin{equation} 
 \label{coupling} 
 V = {1 \over 64\pi^2} \left (M_{\chi}^4 
 \ln \left( {M_{\chi}^2\over 
 \mu^2} \right) - M_{\psi}^4 
 \ln \left( {M_{\psi}^2\over 
 \mu^2} \right) \right )
\end{equation} 
 where $M_{\chi}^2 \equiv \sum _j g_j^2\Phi^{a_j}_j\Phi^{a_j}_j\,,
 ~~ M_{\psi}^2 \equiv \sum _j \tilde{g}_j^2\Phi^{a_j}_j\Phi^{a_j}_j $
 and $\mu$ is a renormalization scale. \\

 This potential induces the coupling among the $i$ and $j$ monopole
 species of the strength $\alpha_{ij} =  g_i^2 g_j^2 - \tilde{g}_i^2 
 \tilde{g}_j^2 \sim g^4$. By a suitable choice of parameters this interaction can be made attractive, repulsive or neutral. 
  Thus, the effective  
 inter-monopole coupling  by the unitarity bound (\ref{Gbound})  scales as 
 $\alpha_{ij} \lesssim 1/N^2$. So the entropy bound formulated as (\ref{Nentropy}) is not violated by the stack of monopoles as long as the system is unitary.  \\
 
 Finally, we notice that at the saturation point $e^2 = g^2 = 1/N$ 
 the area law (\ref{Newbound}) is satisfied. 
 Indeed, the scale $f$ is given by $f^2 = Nv^2$ and we have, 
 \begin{equation} \label{AAA}
   S_{\rm max} = {1 \over \alpha} = N^2 = (Rf)^2  \, .
 \end{equation}
  Thus, although increasing $N$ increases the entropy
  of the stack of monopoles without increasing its size $R$, nevertheless the area 
  measured in units of the order parameter  $f$ increases wth $N$ in such a way that 
  it always matches the entropy at the saturation point.


\begin{thebibliography}{10}

		\bibitem{BekBound}
			
					J.~D. Bekenstein, {\em Universal upper bound on the entropy-to-energy ratio for
				bounded systems\/},  \href{http://dx.doi.org/10.1103/PhysRevD.23.287}{Phys.
				Rev. D {\bf 23} (1981) no.~2, 287--298}. 

			
				
	
	\bibitem{BrBound}
	H.~J. Bremermann, {\em Quantum noise and information\/},  in {\em Proceedings
		of the Fifth Berkeley Symposium on Mathematical Statistics and Probability} {\bf 4} (1967), 15--20.


		\bibitem{BekE}
	J.~D. Bekenstein, {\em Black holes and entropy\/},
	\href{http://dx.doi.org/10.1103/PhysRevD.7.2333}{Phys. Rev. D {\bf 7} (1973)
		no.~8, 2333--2346}.

	
\bibitem{Gia1} G.~Dvali, Area Law Saturation of Entropy Bound from Perturbative Unitarity in Renormalizable Theories,
 arXiv:1906.03530 [hep-th]. 
	

 \bibitem{Mon} G.~'t Hooft, Magnetic monopoles in unified gauge theories, Nucl. Phys. B 79 (1974) 276; \\

 A.M.~Polyakov,  Particle Spectrum in the Quantum Field Theory, 
 JETP Letters,  20 (1974), 194. 


\bibitem{WittenN} 

E.~Witten, Baryons in the 1/n Expansion, 
Nucl. Phys. B160 (1979) 57.


\bibitem{planar} 
G.~'t Hooft,  A Planar Diagram Theory for Strong Interactions", 
Nucl.  Phys.   B72,  (1974) 461. 

\bibitem{Nportrait}
  G.~Dvali and C.~Gomez,
  Black Hole's Quantum N-Portrait,
  Fortsch. Phys.  61 (2013) 742
  [arXiv:1112.3359 [hep-th]].  
  
	G.~Dvali and C.~Gomez, Black Holes as Critical Point of Quantum Phase Transition, Eur. Phys. J. C  74 (2014)  2752, arXiv:1207.4059 [hep-th].	



\bibitem{instanton} A.~Belavin, A.~Polyakov, A.~Schwartz, Y.~Tyupkin, Pseudo-particle solutions of the
Yang-Mills equations, Phys. Lett. B59 (1975) 85.



 \bibitem{scalarinst}
 S. Fubini, 
 A new approach to conformal invariant field theories,
 Nuovo Cimento 34A (1976) 521; \\

L.N. Lipatov, Divergence of the perturbation theory series and 
quasi-classical theory,  Zh. Eksp. Teor. Fiz. 72 (1977) 411; Sov. Phys. JETP
45 (1977) 216.

	
\bibitem{corpuscular}
G.~Dvali, C.~Gomez, L.~Gruending, T.~ Rug, 
Towards a Quantum Theory of Solitons, 
Nucl. Phys.  B901 (2015) 338-353, 
DOI: 10.1016/j.nuclphysb.2015.10.017 
arXiv:1508.03074 [hep-th]; 
Also, the same authors, unpublished. 



\bibitem{review} A.~Vainshtein, V.~Zakharov, V.~Novikov and M.~Shifman, ABC of Instantons, Sov.
Phys. Usp. 25 (1982) 195; Instantons in Gauge Theories, M. Shifman, World Scientific, (Singapore, 1994);
M.~Shifman, A.~Vainshtein, Instanton versus Supersymmetry: Fifteen Years Later,
ITEP Lectures, Edt by M.~Shifman, World Scientific, Singapore 1999, Vol. 2, 485,
hep-th/9902018.

S,~Vandoren, P.~ van Nieuwenhuizen,
Lectures on instantons, arXiv:0802.1862 [hep-th]



\bibitem{Hill}
C.T.~ Hill, P.~ Ramond,  	
Topology in the bulk: Gauge field solitons in extra dimensions, 
 Nucl. Phys. B596 (2001) 243-258 
DOI: 10.1016/S0550-3213(00)00579-4 
hep-th/0007221 

\bibitem{skyrme} 
T.H.R.~Skyrme, A Unified Field Theory of Mesons and Baryons, 
 Nucl. Phys. 31 (1962) 556.  

\bibitem{WittenS}
E. Witten, Global Aspects of Current Algebra, Nucl. Phys. B 223 (1983); 
Current Algebra, Baryons, and Quark Confinement, Nucl.
Phys. B 223 (1983) 433; 
 Skyrmions And Qcd, In *Treiman, S.b. ( Ed.) Et Al.:
Current Algebra and Anomalies*, 529-537 
and Preprint - Witten, E.
(84,REC.OCT.) 7p

 \bibitem{Hol1} 
 
 G.'t~Hooft, Dimensional reduction in quantum gravity, gr-qc/9310026; 
 
 L.~Susskind,
The World As A Hologram, J. Math. Phys. 36, 6377 (1995), hep-th/9409089


\bibitem{speciesG}
	G.~Dvali, \textit{Black Holes and Large N Species Solution to the Hierarchy
		Problem}, Fortschr.\ Phys. {\bf 58} (2010), 528, {\tt arXiv:0706.2050 [hep-th]}.\\
	G.~Dvali and M.~Redi, \textit{Black Hole Bound on the Number of Species and
		Quantum Gravity at LHC},
	Phys.\ Rev.  {\bf D77} (2008),  045027, {\tt arXiv:0710.4344 [hep-th]}.\\
	G.~Dvali and C.~Gomez, \textit{Quantum Information and Gravity Cutoff in Theories
		with Species}, Phys.\ Lett.  {\bf B674} (2009),  303,
	{\tt arXiv:0812.1940 [hep-th]}.
 

	
\bibitem{DNT}
G.~Dvali, H.B.~Nielsen, N.~Tetradis, 
Localization of gauge fields and monopole tunnelling 
 Phys.Rev. D77 (2008) 085005 
DOI: 10.1103/PhysRevD.77.085005,  
arXiv:0710.5051 [hep-th]

\bibitem{prefactor}

Yu.A.~Bashilov and S.V.~Pokrovsky, Nucl. Phys. B143 (1978)
431. 

C.~Bernard, Gauge zero modes, instanton determinants, and quantum-chromodynamics calculations, Phys. Rev. D 19 (1979) 3014. 


\bibitem{GW}  D.J.~Gross and E.~Witten,  
Possible third order phase transition in the large-$N$ lattice gauge theory, 
Phys. Rev. D 21, (1980) 446.  

\bibitem{Wadia} 
 S.R.~Wadia, A Study of $U(N)$ Lattice Gauge Theory in 2-dimensions, arXiv:1212.2906 [hep-th], an edited version of the 
unpublished 1979 preprint, EFI-79/44-CHIC

\bibitem{Neu}
H.~Neuberger, Nonperturbative Contributions in Models With a Nonanalytic Behavior at Infinite N,
Nucl. Phys. B 179, 253 (1981).


\bibitem{species} D.~Marolf, D.~Minic, and S.F.~Ross, Notes on
space-time thermodynamics and the observer
dependence of entropy, Phys.Rev. D69 (2004) 064006,
arXiv:hep-th/0310022 [hep-th].


\bibitem{B1} R.~Bousso, Bound states and the Bekenstein bound, 
	JHEP 0402 (2004) 025, 	
DOI:	10.1088/1126-6708/2004/02/025, 
	arXiv:hep-th/0310148

\end{thebibliography}
\end{document}